\documentclass[a4paper,11pt]{article}
\usepackage{jheppub}

\usepackage{graphicx}
\usepackage{dcolumn}
\usepackage{bm}
\usepackage[title]{appendix}
\usepackage{pstricks}
\usepackage{color}
\usepackage{hyperref}
\hypersetup{
     colorlinks   = true,
     linkcolor    = red,
     citecolor    = blue
}

\usepackage{cleveref}
\crefname{table}{Table}{Tables}
\crefname{equation}{Eq.}{Eqs.}
\crefname{appendix}{App.}{Apps.}
\crefname{section}{Sec.}{Secs.}
\crefname{figure}{Fig.}{Figs.}
\usepackage{xcolor}
\usepackage{graphicx}
\usepackage{braket}
\usepackage{tikz}
\usetikzlibrary{decorations, decorations.markings, decorations.pathmorphing, arrows, graphs, shapes.geometric, snakes}
\tikzset{
    photon/.style={
        decoration={snake, amplitude=0.15cm, segment length=0.2cm},
        decorate    
    },
    fermion/.style={
        decoration={
            markings,
            mark=at position 0.5 with {\node[transform shape, xshift=-0.5mm, fill=black, inner sep=1pt, draw, isosceles triangle]{};}
        },
        postaction=decorate
    },
    scalar/.style={
        dashed,
        decoration={
            markings,
            mark=at position 0.5 with {\node[transform shape, xshift=-0.5mm, fill=black, inner sep=1pt, draw, isosceles triangle]{};}
        },
        postaction=decorate
    },
    gluon/.style={
        decoration={coil, aspect=0.75, mirror, segment length=1.5mm},
        decorate
    }, 
    left/.style={
        bend left=90,
        looseness=1.75
    },
    higgs/.style={
      thick,dashed
    },
    loryon/.style={
      thick
    }
}

\usepackage{mathtools}

\usepackage{svg}

\usepackage{bbold}

\usepackage{booktabs}
\usepackage{multirow}

\DeclareMathOperator{\Tr}{Tr}

\newcommand{\lag}{\mathcal{L}}

\newcommand{\abs}[1]{\left| #1 \right|}
\newcommand{\dd}{\mathrm{d}}

\newcommand{\mK}{\mathcal{K}}
\newcommand{\mM}{\mathcal{M}}
\newcommand{\mT}{\mathcal{T}}

\title{\boldmath Non-decoupling scalars at future colliders}

\author[a]{Graeme Crawford,}
\author[a]{Dave Sutherland}
\affiliation{School of Physics and Astronomy, University of Glasgow, Glasgow G12 8QQ, UK}

\emailAdd{g.crawford.2@research.gla.ac.uk}
\emailAdd{david.w.sutherland@glasgow.ac.uk}

\abstract{
We consider a class of BSM models where a generic scalar electroweak multiplet obtains a significant fraction of its mass from a coupling to the Higgs. Such models are non-decoupling: their new states are necessarily at the TeV scale or below, they can significantly alter the electroweak phase transition, and they have a pattern of low energy effects that are distinct from those predicted by SMEFT.
Using their minimal gauge and Higgs couplings, we show that a future precision lepton collider (such as FCC-ee, CEPC, ILC, or CLIC) can probe all the non-decoupling parameter space of scalar electroweak multiplets, providing fundamental information on the mechanism of electroweak symmetry breaking.}

\begin{document}
\maketitle
\flushbottom

\section{Introduction}
\label{sec:intro}

Despite increasingly precise measurements of the Higgs boson, a question remains over the nature of electroweak symmetry breaking (EWSB). Observations of the Higgs at the LHC are made in the zero temperature phase where electroweak symmetry is broken, and where single Higgs couplings to other particles are thus far consistent with the Standard Model (SM) expectation \cite{ATLAS:2022vkf,CMS:2022dwd}. However, we have not experimentally tested the full potential landscape nor the phase diagram of the underlying theory. To do this requires future experiments to measure higher Higgs multiplicities and finite temperature phenomena in the early Universe \cite{abada_fcc_2019,Colpi:2024xhw}.

If the nature of electroweak symmetry breaking is altered significantly from the SM prediction, then the beyond the Standard Model (BSM) physics responsible is by definition \emph{non-decoupling}, in the sense that its associated mass scale cannot be made arbitrarily high. Roughly speaking, new physics that significantly distorts the dynamics of the SM path of length $v$ from electroweak preserving vacuum to electroweak breaking vacuum is associated with a mass scale of $4 \pi v$ or below, by na\"ive dimensional analysis. Are any such TeV-scale models of non-decoupling physics still viable?

Non-decoupling physics divides usefully into two categories. The first category is models which contain significant new sources of electroweak symmetry breaking that contribute to the $W$ and $Z$ masses, such as an extended scalar sector or some technicolor-like chiral condensate.\footnote{Note that modern composite Higgs models do not fall into this category, as the symmetry breaking pattern of the strong dynamics (associated with the order parameter $f$) leaves electroweak symmetry intact. Its decoupling behaviour is manifest when taking $f \to \infty$.} The second category is new particles that get all or most of their mass from coupling to the Higgs, and which therefore cannot be taken to be arbitrarily heavy without violating unitarity bounds on that coupling.\footnote{There is a sense in which those in the first category also belong to the second category. For example, new states in generic extended scalar sectors contribute to $W$ and $Z$ masses through their own vev, \emph{and} their own mass is also dependent on the Higgs vev.} Both such categories of non-decoupling models can cause a strongly first order electroweak phase transition in the early universe (which is a necessary condition for a theory of electroweak baryogenesis that explains the observed matter-antimatter asymmetry in the Universe) \cite{Caldwell:2022qsj,Asadi:2022njl}. Both categories also predict a pattern of deviations from the SM in low energy observables, which is distinct from the pattern predicted within the power counting of SMEFT \cite{Cohen:2020xca}. Both categories also have a finite parameter space to explore, as they must be associated with TeV-scale new states. Non-decoupling new physics can therefore be definitively ruled in or ruled out, with either result giving fundamental information on the mechanism of electroweak symmetry breaking. It presents a finite and well-motivated target for experiments.

The two categories of non-decoupling physics have distinct Higgs phenomenology. Generic models containing extra sources of EWSB, and which therefore contribute to the $W$ and $Z$ masses, predict large deviations in single Higgs couplings, such as $hWW$ and $hZZ$. For instance, in a weakly coupled extended scalar sector the deviations result from tree-level mixing between an SM-like Higgs and other neutral scalar components; such non-decoupling models are now constrained to relatively small regions of parameter space where the tree-level mixing is suppressed, yet there remain sizeable vevs in scalar components other than the SM Higgs doublet.

By contrast, the effects of a new particle gaining most of its mass through a coupling to the Higgs can be much more minimal. This may be a fermion or a scalar with a largely unbroken $\mathbb{Z}_2$ symmetry; this symmetry prevents the scalars falling into the less minimal `extra sources of EWSB' class. As the entire Standard Model is built from particles gaining their mass from the Higgs, it is natural to ask if there are any more beyond the Standard Model. (\cite{banta_non-decoupling_2022} used the term \emph{Loryon} for any particle gaining most of its mass through a coupling to the Higgs, which we will also use as a convenient shorthand in this paper.) Regarding their experimental signatures, any Loryons with SM gauge charges contribute through loops to $h\gamma\gamma$ or $hgg$ couplings. Measurements of $\kappa_\gamma$ and $\kappa_g$ at the time of the Higgs discovery, together with electroweak precision observables and direct searches, were sufficient to rule out an entire chiral 4\textsuperscript{th} generation of quarks and leptons \cite{Djouadi:2012ae,Kuflik:2012ai}. A decade later, a more general analysis of Loryons organised by their SM gauge charges, shows that more possibilities can be ruled out \cite{banta_non-decoupling_2022}. Absent large cancellations between different BSM multiplets, all Loryons with colour charge will be detectable by the end of the HL-LHC run, as well as all fermions. However, despite their non-decoupling nature, HL-LHC may not be sufficiently sensitive to detect a handful of small electroweak scalar irreps, therefore leaving some Loryon possibilities open.

This paper aims to show that a future $e^+e^-$ collider can definitively answer if any natural model of electroweakly charged non-decoupling physics is present. We use the projected sensitivities for FCC-ee as a benchmark in this paper; however, the same conclusions hold for machines such as CEPC, ILC, and CLIC. We show that the precision measurement of order $10^{12}$ $Z$s, $10^8$ $W$ pairs, and crucially $10^6$ $h$s, is almost exactly what's required to detect the remaining non-decoupling scalar possibilities. In so doing, we explore a higher mass range than the similar-spirited works of \cite{AbdusSalam:2013eya,Katz:2014bha}.

In \cref{sec:loryon} we state the minimal model of a $\mathbb{Z}_2$-symmetric scalar electroweak multiplet in terms of its Higgs and gauge couplings, focussing on its non-decoupling region of parameter space where Higgs couplings are large. This model encapsulates the minimal loop-level, indirect effects of many BSM models of new physics, including models where the scalars break the $\mathbb{Z}_2$ symmetry, but have their parameters tuned to remove tree-level effects. \cref{sec:unitarity} shows that the states in the multiplet must have sub-TeV masses by unitarity; \cref{sec:ewpo} shows that their viable mass range is probed from above and below by FCC-ee's Higgs and $WW/Z$ runs respectively --- the latter also put a strong bound on the mass splittings between different components in the same multiplet. \cref{sec:baryogenesis} shows that these precision measurements are sufficient to probe the region of phase space that causes a strongly first order electroweak phase transition in the early Universe. We argue in \cref{sec:discussion} that the sum of these measurements constitutes a rather definitive probe of non-decoupling new physics.

\section{The Loryon model}
\label{sec:loryon}

Consider an arbitrary scalar electroweak multiplet, $\Phi$, in a dimension $2j+1$ irrep of $SU(2)_L$ and with hypercharge $Y$ under $U(1)_Y$. We impose a $\mathbb{Z}_2$ symmetry to forbid tree-level mixing of its neutral components with the physical Higgs, and in the following we study only the universal, loop-level effects of the multiplet $\Phi$. We assume that the parameters of the scalar potential are such that $\Phi$ does not acquire a vev, and the $\mathbb{Z}_2$ is never spontaneously broken, even at finite temperature.

We note that the $\mathbb{Z}_2$ symmetry must be \emph{explicitly} broken by a small amount on phenomenological grounds, in order to allow the charged components of $\Phi$ to decay promptly, and thereby avoid creating exotic signatures such as disappearing tracks in colliders \cite{ATLAS:2018imb,CMS:2016kce}. However, the degree of breaking required is small and technically natural. Depending on the decay products of the charged components, current constraints from direct searches for these decays do not bound their masses above a few hundred GeV \cite{banta_non-decoupling_2022}. For example, a singly charged scalar of mass $\sim 350 \text{ GeV}$ can decay undetected into a charged lepton and neutrino \cite{Crivellin:2020klg}; the charged components of an inert second Higgs doublet can decay electroweakly to neutral ones, which escape the detector, leaving the leading bound on the charged particles at around $\sim 70 \text{ GeV}$ \cite{Dercks:2018wch}, and a real triplet of mass $\sim 230 \text{ GeV}$ can decay by a small mixing with the Higgs without detection \cite{Bell:2020gug}. The real scalar singlet, unless it is sufficiently light to be pair produced in Higgs decays, is currently effectively unconstrained \cite{Carena:2019une}; nor will measurements of its off-shell Higgs production at HL-LHC reach above a couple of hundred GeV \cite{Craig:2014lda}.

In short, much of the parameter space we consider below is not presently ruled out. In some expressions, we will take a benchmark mass for the Loryons of $600$ GeV, which can be assumed to be safe from the present interpretation of direct search constraints given in \cite{banta_non-decoupling_2022}. We therefore turn to universal, indirect bounds instead.

The above assumptions lead to the following class of simplified models, which are intended to span the minimal loop-level indirect effects of all scalars with a significant coupling to the Higgs. A general $\mathbb{Z}_2$-symmetric electroweak multiplet $\Phi$ will have gauge and Higgs couplings, these latter through up to three distinct cross-quartic interactions between $\Phi$ and the Higgs doublet $H$. For a complex multiplet, the general renormalisable Lagrangian is
\begin{align}
    \lag =& \abs{D \Phi}^2 - V(\Phi) \, , \\
    V(\Phi) =& \left( m_\text{ex}^2  + \lambda \abs{H}^2 \right) \abs{\Phi}^2 \notag\\
    & \hspace{1ex} + \lambda^\prime ( \Phi^\dagger T^a \Phi ) ( H^\dagger T^a H ) + \begin{cases}
        \lambda^{\prime\prime} ( \tilde{\Phi}^\dagger T^a \Phi ) ( H^\dagger T^a \tilde{H} ) + \text{h.c.} & \text{ if } Y=+\frac12 \\
        \lambda^{\prime\prime} ( \Phi^\dagger T^a \tilde{\Phi} ) ( H^\dagger T^a \tilde{H} ) + \text{h.c.} & \text{ if } Y=-\frac12 \\
        0 & \text{ otherwise}
    \end{cases} \, , \label{eq:complexPot}
\end{align}
where $D = \partial - i g W^a T^a - i {g^\prime} B Y$.

The multiplet $\Phi$ has an explicit mass term $m_\text{ex}^2$, and upon electroweak symmetry breaking its components receive additional mass contributions through their cross-quartic interactions with the Higgs doublet $H$. Of these, $\lambda$ generates a uniform mass shift, whereas $\lambda^\prime$ and $\lambda^{\prime\prime}$ split the masses of the components. The $\lambda^{\prime\prime}$ term appears only for multiplets with $\abs{Y}=\frac12$, and is expressed in terms of
\begin{equation}
    \tilde{\Phi} = S \Phi^* \, .
\end{equation}
$S$ is a similarity transformation that maps a representation of $SU(2)$ onto its equivalent conjugate representation. In the simplest example of such a $\abs{Y}=\frac12$ multiplet --- the extra doublet of the inert two Higgs Doublet Model --- the couplings correspond to $\lambda_3 = \lambda - \frac{\lambda^{\prime}}{4}$, $\lambda_4 = \frac{\lambda^{\prime}}{2}$, and $\lambda_5 = \lambda^{\prime\prime}$ in the more familiar notation (see, e.g., \cite{Davidson:2005cw}).

Equivalently, for a real multiplet with $Y=0$, the general renormalisable Lagrangian is
\begin{align}
    \lag =& \frac12 \left(D \Phi\right)^2 - V(\Phi) \, , \\
    V(\Phi) =& \frac12 \left( m_\text{ex}^2  + \lambda \abs{H}^2 \right) \Phi^2 + \frac12 i \lambda^\prime ( \Phi T_R^a \Phi ) ( H^\dagger T^a H ) \, , \label{eq:realPot}
\end{align}
where $D = \partial + g W^a T^a_R$, in terms of the real generators $T^a_R$. We present our conventions for $S$ and the $SU(2)$ generators $T^a$, $T^a_R$ in \cref{app:su2conventions}.

\subsection{Mass Spectrum}
\label{subsec:spectrum}

The mass spectrum for a generic multiplet is a function of $\{m_\text{ex}^2,\lambda,\lambda^\prime,\lambda^{\prime\prime}\}$, as seen by writing the Higgs doublet in unitary gauge
\begin{equation}
    H = \frac{1}{\sqrt{2}} \begin{pmatrix}
        0 \\ v+h
    \end{pmatrix} \, ,
\end{equation}
and substituting into \cref{eq:complexPot,eq:realPot}. We identify an overall mass scale for the new states
\begin{equation}
    M^2 = m_\text{ex}^2 + \frac12 \lambda v^2 \, .
    \label{eq:overallMassScale}
\end{equation}
It is useful to then define three dimensionless quantities to describe the remaining parameter freedom: the fraction $f$ of $M^2$ that is due to the coupling $\lambda$ with the Higgs, together with two coefficients that parametrise the degree of mass splitting due to couplings $\lambda^\prime, \lambda^{\prime\prime}$ with the Higgs,
\begin{equation}
    f = \frac{\lambda v^2}{2 M^2} \, ; \qquad r_1 = \frac{\lambda^\prime v^2}{4 M^2} \, ; \qquad r_2 = \frac{\lambda^{\prime\prime} v^2}{4 M^2} \, .
\label{eq:notation}
\end{equation}
$r_1$ controls the mass splitting between components of different weak isospin. The effects of the $r_2$ term is to mix and split the masses of components with equal or opposite electric charge.\footnote{In the simplest case of the inert 2HDM, $r_2$ splits the masses of the neutral scalar and pseudoscalar components, usually denoted $H$ and $A$ respectively. For generic irreps, $r_2$ causes mixing as well.} \cref{tab:modelMassSpectra} gives the mass spectra of the simplest irreps. For a generic irrep, the mass spectrum and single Higgs couplings of the Loryons are extracted from the potential
\begin{equation}
    \label{eq:scalarPotential}
    V(\Phi) = \left(1 + h \frac{\partial}{\partial v} \right)M^2 \times \begin{cases}
        \Phi^\dagger \left(\mathbb{1}-r_1 T^3\right) \Phi  + \left( \frac12 \Phi^T N \Phi + \text{h.c.} \right) & \text{ complex } \\
        \frac12 \Phi^T \left(\mathbb{1}-i r_1 T^3_R\right) \Phi & \text{ real}
    \end{cases} \, ,
\end{equation}
where
\begin{equation}
    \label{eq:secondMassSplitting}
    N = \begin{cases}
         -2 r_2  S^\dagger T^+ & \text{ if } Y=+\frac12 \\
        -2 r_2^* S^\dagger T^- & \text{ if } Y=-\frac12 \\
        0 & \text{ otherwise}
    \end{cases} \, .
\end{equation}

\begin{table}
  
  \centering
  \begin{tabular}{cccc}
    & \textbf{Singlet}&\textbf{Real triplet}&\textbf{2HDM}\\
  \cmidrule{2-4}\morecmidrules\cmidrule{2-4} && \\[-1ex]
  $\mathbf{\Phi}$ & $\begin{pmatrix}
    s
  \end{pmatrix}$ & $\begin{pmatrix}
    H^+ \\ H^0 \\ H^-
  \end{pmatrix}$ & $\begin{pmatrix}
    H^+ \\ \frac{1}{\sqrt{2}} \left(H + i A\right)
  \end{pmatrix}$ \\ \addlinespace[0.5em] \cmidrule{2-4} \addlinespace[0.5em]
  \textbf{Masses} & 
  \parbox{23mm}{\begin{equation*} m_s^2  = M^2  \,\, {\color{gray}\{1\}} \end{equation*}\vspace*{-5pt}} & 
  \parbox{41mm}{\begin{align*}
    m_{H^{+}}^2 =& \, M^2 \left(1- r_1\right) \,\, {\color{gray}\{1\}}  \\
    m_{H^{0}}^2 =& \, M^2  \,\, {\color{gray}\{1\}} \\
    m_{H^{-}}^2 =& \, M^2 \left(1- r_1\right) \,\, {\color{gray}\{1\}} 
\end{align*}\vspace*{-4pt}} &
\parbox{56mm}{\begin{align*}
  m_{H^\pm}^2 =& \, M^2 \left(1-\frac12 r_1\right)  \,\, {\color{gray}\{2\}} \\
  m_{H}^2  =& \, M^2 \left(1+\frac12 r_1+2r_2\right)  \,\, {\color{gray}\{1\}} \\
  m_{A}^2 =& \, M^2 \left(1+\frac12 r_1-2r_2\right)  \,\, {\color{gray}\{1\}} 
\end{align*}\vspace*{-2pt}} 
  \end{tabular}
\caption{The mass spectrum for the scalar representations considered in this work, presented in terms of the parameters of \cref{eq:notation}. The number of real degrees of freedom associated to each mode is shown in grey braces. Note that the mass spectrum for the complex triplet (where the numbers of real degrees of freedom are doubled) takes the same form as the real triplet, however the charge of each component is shifted by any hypercharge carried by the triplet.
\label{tab:modelMassSpectra}}
\end{table}

If any of the dimensionless parameters $f$, $r_1$, and $r_2$ are order one, then some components of the multiplet obtain an order one fraction of their mass from coupling to the Higgs. This gives the particles non-decoupling behaviour, such as being significantly lighter when the Higgs vev is turned off. For any mass eigenstate of the multiplet, we can define the fraction of its mass squared $m_i^2$ coming from its coupling to the Higgs
\begin{equation}
  f_i = \frac{\partial \log m_i^2}{\partial \log v^2} \, .
  \label{eq:fidef}
\end{equation}
Note that, in the absence of mass splitting, $m_i^2=M^2$ and $f_i=f$ for all components. If any component has $f_i \geq \frac12$, such that it gets more than half of its mass from the Higgs, the SMEFT expansion for this model does not converge at our vacuum. Consequently, their low energy behaviour is poorly approximated by matching on to SMEFT \cite{Cohen:2020xca}.

As we will show, order one values of $f$, $r_1$, and $r_2$ (which match on to large values of $\lambda$, $\lambda^\prime$, and $\lambda^{\prime\prime}$ respectively) are significantly constrained by measurements at a future $e^+ e^-$ machine.

\subsection{Generic Lagrangian\label{sec:genericLag}}

Here we define some notation that is useful for the intermediate calculational steps in this paper.

For the purposes of treating both complex and real multiplets equivalently in the following calculations, as well as accounting for the mixing effects of $r_2$, we write all Lagrangians in the common form
\begin{align}
  \lag =& \frac12 (\partial F)^T \mK (\partial F) - \frac12 \left(1 + h \frac{\partial}{\partial v} \right) F^T \mM F \notag\\
  & \hspace{2ex} - \sum_X \frac12 g_X X^\mu \, F^T \mK \mT^X \overset{\leftrightarrow}{\partial_\mu} F - \sum_{X,Y} \frac12 g_X g_Y X^\mu Y_\mu \, F^T \mK \left\{ \mT^X, \mT^Y \right\} F \, ,\label{eq:genLag}
\end{align}
where the gauge bosons $X,Y$ and respective couplings $g_X,g_Y$ are drawn from
\begin{align}
    X,Y &\in \{ W^+, W^-, W^3, B \} \, , \\
    g_X,g_Y &\in \{ \frac{g}{\sqrt{2}}, \frac{g}{\sqrt{2}}, g, g^\prime \} \, .
\end{align}
For a complex multiplet, the generic Lagrangian \cref{eq:genLag} is built from $2(2j+1)$-by-$2(2j+1)$ matrices, given in block form by
\begin{gather}
  F = \begin{pmatrix}
    \Phi \\ \Phi^*
  \end{pmatrix} \, ; \,\,
  \mK = \begin{pmatrix}
    0 & \mathbb{1} \\ \mathbb{1} & 0
  \end{pmatrix} \, ; \notag\\[5pt]
  \mM = M^2 \begin{pmatrix}
    N & \left(\mathbb{1}-r_1 T^3\right)^T \\ \mathbb{1}-r_1 T^3 & N^*
  \end{pmatrix} \, ; \,\,
  \mT^A = \begin{pmatrix}
    -i T^A & 0 \\ 0 & i \left(T^A\right)^T
  \end{pmatrix} \, . \label{eq:complexCals}
\end{gather}
For the real case, it is built from the $(2j+1)$-by-$(2j+1)$ matrices given by
\begin{equation}
  F = \Phi \, ; \,\,
  \mK = \mathbb{1} \, ; \,\,
  \mM = M^2 \left(\mathbb{1}-i r_1 T^3_R\right) \, ; \,\,
  \mT^A = T^A_R \, . \label{eq:realCals}
\end{equation}
Note that $\mK = \mK^{-1} = \mK^T$ and $\mK \mT^A = - \left(\mT^A\right)^T \mK$ in both the real and complex cases.

\section{Unitarity Constraints}
\label{sec:unitarity}

We are interested in particles which obtain a significant fraction of their mass from a necessarily large coupling to the Higgs. The size of such a coupling is bounded from above by perturbative unitarity, and it follows that the masses of these non-decoupling particles are bounded too; the new states must reside at or below the TeV scale, as we now demonstrate.

We consider the constraints arising from elastic 2-to-2 scalar scattering at tree level. These are dominated by the elastic scattering of the new scalar multiplet with itself, via the exchange of a Higgs \cite{banta_non-decoupling_2022}. This is because it is the only tree diagram to grow as the square of the necessarily large coupling to the Higgs.

Let the real field $\phi_i$ be a mass-eigenstate combination of the components of the multiplet $F$, with mass $m_i^2$ the corresponding eigenvalue of $\mM$, of which a fraction $f_i$ comes from EWSB. The Higgs-exchange amplitude between $\phi_i$ and itself is
\begin{align}
  \mathcal{A}\left(\phi_i \phi_i \to \phi_i \phi_i\right) =& -\left(\frac{2 f_i m_i^2}{v}\right)^2 \left(\frac{1}{t - m_h^2} + \frac{1}{u - m_h^2}\right) \, . \label{eq:higgsExchangeAmp}
\end{align}
Recall that, in the absence of mass splitting, $m_i^2=M^2$ and $f_i=f$ for all components.

We project out the $s$-wave component of the amplitude using the expressions in \cite{Goodsell:2018tti}
\begin{align}
  a_0(s) =& \frac{1}{64 \pi} s^{-\frac12} \left(s-4 m_i^2\right)^{\frac12} \int_{-1}^1 \dd(\cos \theta) \, \mathcal{A}(s,\theta) \, , \notag\\
  =& \frac{1}{4 \pi} \left(\frac{f_i m_i^2}{v}\right)^2 s^{-\frac12} \left(s-4 m_i^2\right)^{-\frac12} \log \left(1 + \frac{s-4m_i^2}{m_h^2}\right) \, , \label{eq:a0s}
\end{align}
whose real part is constrained by perturbative unitarity to be less than a half for all kinematically accessible values of $s$:
\begin{equation}
  \max_{4m_i^2 \leq s < \infty} \Re a_0(s) \leq \frac12 \, . \label{eq:unitarityConditiona0}
\end{equation}
When $\frac{m_h^2}{4 m_i^2} \ll 1$, $a_0(s)$  peaks at a few $m_h^2$ above the threshold value of $s=4 m_i^2$. Numerically maximising the real part of \cref{eq:a0s} reduces \cref{eq:unitarityConditiona0} to the constraint
\begin{align}
  \frac{1}{8 \pi} \frac{m_i^3 f_i^2}{v^2 m_h} \left( 0.81 - 1.29 \frac{m_h^2}{4 m_i^2} + \mathrm{O}\left(\frac{m_h^2}{4 m_i^2}\right)^2 \right) \leq \frac12 \, . \label{eq:unitarityCond}
\end{align}
At leading order in $\frac{m_h^2}{4 m_i^2}$ this reduces to
\begin{equation}
  \left( \frac{m_i}{600 \, \mathrm{GeV}} \right)^3 f_i^2 \leq 0.54 \, .
  \label{eq:approxUnitarityCond}
\end{equation}
Thus, if $f_i$ is order one for any state, then it must have a mass below $1$ TeV.

Note that the fuller analysis in \cite{banta_non-decoupling_2022} yields mass bounds a few tens of GeV different --- notably this includes the scenario where a contact piece is added to \cref{eq:higgsExchangeAmp} arising from a quartic self-interaction of the new multiplet. In any case, we stress that perturbative unitarity yields only approximate bounds on the validity of the theory, and \cref{eq:approxUnitarityCond} should be interpreted as such.

\begin{table}
  \centering
  \begin{tabular}{c c c c}
    \textbf{Observable} & \textbf{Representative diagram} & \textbf{Scaling} & \textbf{Expression} \\ \hline\hline \addlinespace[0.5em]
    Unitarity &
    \begin{tikzpicture}[baseline=-0.65ex]
      \coordinate (v1) at (0,0.6);
      \coordinate (v2) at (0,-0.6);
      \draw[loryon] (-1,1) -- (v1) -- (1,1);
      \draw[loryon] (-1,-1) -- (v2) -- (1,-1);
      \draw[higgs] (v1) -- (v2);
    \end{tikzpicture} &
    $M^3 f^2$ & \cref{eq:unitarityCond} \\\addlinespace[0.5em] \hline \addlinespace[0.5em]
    $S$ &
    \multirow{4}{*}{\begin{tikzpicture}[scale=0.4,baseline=-0.65ex]
      \draw[loryon] (0,0) arc (0:180:1);
      \draw[loryon] (-2,0) arc (180:360:1);
      \draw[photon] (-3.5,0) -- (-2,0);
      \draw[photon] (0,0) -- (1.5,0);
    \end{tikzpicture}} &
    $r_1 Y C(j)$ or $r_2^2 Y C(j)$ & \cref{eq:generalS} \\ \addlinespace[0.3em]
    $T$ & &
    $M^2 r_1^2 C(j)$ or $M^2 r_2^2 C(j)$ & \cref{eq:generalT} \\ \addlinespace[0.3em]
    $W$ & &
    $M^{-2} C(j)$ & \cref{eq:generalW} \\ \addlinespace[0.3em]
    $Y$ & &
    $M^{-2} Y^2 d(j)$ & \cref{eq:generalY} \\ \addlinespace[0.5em] \hline \addlinespace[0.5em]
    $\kappa_h$ &
    \begin{tikzpicture}[scale=0.4,baseline=-0.65ex]
      \draw[loryon] (0,0) arc (0:180:1);
      \draw[loryon] (-2,0) arc (180:360:1);
      \draw[higgs] (-3.5,0) -- (-2,0);
      \draw[higgs] (0,0) -- (1.5,0);
    \end{tikzpicture} &
    $M^2 f^2 d(j)$ & \cref{eq:kappahCorrection} \\ \addlinespace[0.5em] \hline \addlinespace[0.5em]
    $\kappa_\gamma$ &
    \begin{tikzpicture}[scale=0.9,baseline=-0.65ex]
      \coordinate (v1) at (0,0);
      \coordinate (v2) at (0.87,0.5);
      \coordinate (v3) at (0.87,-0.5);
      \draw[higgs] (v1) -- +(-0.7,0);
      \draw[photon] (v2) -- +(0.7,0);
      \draw[photon] (v3) -- +(0.7,0);
      \draw[loryon] (v1) -- (v2) -- (v3) -- cycle;
    \end{tikzpicture} &
    $f \left(C(j) + Y^2 d(j)\right)$ & \cref{eq:kappaGamma} \\ \addlinespace[0.5em] \hline \addlinespace[0.5em]
    $\kappa_\lambda$ &
    \begin{tikzpicture}[scale=0.9,baseline=-0.65ex]
      \coordinate (v1) at (0,0);
      \coordinate (v2) at (0.87,0.5);
      \coordinate (v3) at (0.87,-0.5);
      \draw[higgs] (v1) -- +(-0.7,0);
      \draw[higgs] (v2) -- +(0.7,0);
      \draw[higgs] (v3) -- +(0.7,0);
      \draw[loryon] (v1) -- (v2) -- (v3) -- cycle;
    \end{tikzpicture} &
    $M^4 f^3 d(j)$ & \cref{eq:kappalambda} \\
  \end{tabular} 
  \caption{The unitarity constraints considered in \cref{sec:unitarity}, and the observables considered in \cref{sec:ewpo}, together with a representative diagram and the scaling of the Loryon's contribution. The Loryons are shown as solid lines and the Higgs as dashed lines. The Loryon contribution's scaling is given in terms of the mass and coupling parameters $\{M^2,f,r_1,r_2\}$ of \cref{eq:overallMassScale,eq:notation}, the gauge parameters of hypercharge $Y$ and $C(j)=\Tr [(T^3)^2]=\frac23 j (j+\frac12) (j+1)$, and the dimension of the irrep $d(j)=2j+1$. The scaling is stated in the limit of small mass splitting, i.e., to leading order in $r_1$ and $r_2$, and also neglects the masses of the SM states. \label{tab:obs}}
\end{table}

\section{Electroweak and Higgs measurements}
\label{sec:ewpo}

In this section we consider the key indirect effects of a $\mathbb{Z}_2$-symmetric multiplet in precision observables, which arise from its gauge and Higgs couplings. Such effects are naturally loop level, and in the interest of universality, we focus on the oblique observables that are independent of the multiplet's couplings to SM fermions.

We calculate contributions to $S$, $T$, $W$, and $Y$, which parametrise the leading deviations in the electroweak gauge bosons' two-point functions \cite{barbieri_electroweak_2004}, as well as $\kappa_h$, which acts similarly for the Higgs two-point function. All of these would be measured to order-of-magnitudes better precision during the $Z$, $WW$, and $h$ runs of a collider such as FCC-ee. We also consider the loop-level contributions to $\kappa_\gamma$ (the deviation in the $h\gamma\gamma$ coupling) and $\kappa_\lambda$ (the deviation in the $hhh$ coupling), which would be measured to greater precision at FCC-ee and FCC-hh respectively.

\cref{tab:obs} summarises the relevant observables.

\subsection{Oblique parameters: $S$, $T$, $W$, $Y$\label{subsec:zwwrun}}

All electroweakly-charged multiplets give universal loop-level corrections to gauge boson self-energies, which can be characterised by the oblique parameters. The most sensitive oblique parameters are $S$ and $T$, unless new physics carries very large $SU(2)_L$ representations \cite{Lavoura:1993nq}. Note that $U$ is suppressed by a factor $\frac{m_W}{M}$ relative to $T$ \cite{barbieri_electroweak_2004}. Here we present general expressions for the shift in $S$, $T$, $W$, and $Y$ due to any irrep in terms of the mass splitting parameters defined in \cref{eq:notation}. We do so in the $m_Z^2 \to 0$ limit, which we show to be a good approximation.

Where the vacuum polarization for the gauge bosons is defined in the mostly minus metric as a momentum dependent interaction
\begin{equation}
  \lag = \frac12 \Pi_{AB}(p^2) A_\mu(p) B^\mu(-p) \, ,
\end{equation}
the one-loop dimensionally regularised contributions from an additional electroweak scalar multiplet, $\delta \Pi_{AB}$, are given at zero momentum by
\begin{align}
  \frac{2 (4\pi)^\frac{d}{2}}{g_A g_B} \delta \Pi_{AB}(0) =& \frac{1}{\Gamma\left({\textstyle \frac{d}{2}}+1\right)} \int \dd t \, t^{\frac{d}{2}} \Tr \left[ \left[\Delta^{-1} ,\mT^B \right] \Delta^{-1} \mT^A \right] + \left\{ A \leftrightarrow B \right\} \, , \label{eq:genTintSimplified} \\
  \frac{(4\pi)^\frac{d}{2}}{g_A g_B} \delta \Pi^\prime_{AB}(0) =& \frac{1}{\Gamma\left({\textstyle \frac{d}{2}}+2\right)} \int \dd t \, t^{\frac{d}{2}+1} \Tr \left[ \Delta^{-2} \mT^B \Delta^{-2} \mT^A \right] \, ,  \label{eq:genSintSimplified} \\
  \frac{2 (4\pi)^\frac{d}{2}}{g_A g_B} \delta \Pi^{\prime\prime}_{AB}(0) =& \frac{1}{\Gamma\left({\textstyle \frac{d}{2}}+3\right)} \int \dd t \, t^{\frac{d}{2}+2} \Tr \big[ 67 \Delta^{-3}\mT^B \Delta^{-3} \mT^A - 78 \Delta^{-4}\mT^B \Delta^{-2} \mT^A \notag\\
  &\hspace{23ex} + 54 \Delta^{-5}\mT^B \Delta^{-1} \mT^A \big] + \left\{ A \leftrightarrow B \right\} \, , \label{eq:genWintSimplified}
\end{align}
in terms of the quantities \cref{eq:complexCals,eq:realCals}, where $\Delta = t \mathbb{1} + \mK^{-1} \mM$ and $t = k^2$ is the square of the Euclideanised loop momentum. To evaluate these integrals in the limit of small mass splitting, we expand the mass matrix about the identity
\begin{equation}
     \mK^{-1} \mM = M^2 \mathbb{1} + \mK^{-1} \delta\mM \, ,
\end{equation}
where $\mK^{-1} \delta\mM$ can be expressed in terms of mass splitting parameters $r_1$ and $r_2$ as per \cref{eq:complexCals,eq:realCals}.

The contributions of an arbitrary multiplet to $S$, $T$, $W$, and $Y$ are finite, and can be written in the $m_Z^2 \to 0$ limit as \cite{barbieri_electroweak_2004}\footnote{Note the overall difference of $-\frac{1}{g_A g_B}$ in our normalisation of $\Pi_{AB}$ compared to \cite{barbieri_electroweak_2004}, which uses the mostly plus metric and non-canonically normalised gauge bosons.}
\begin{align}
  \Delta S =& -\frac{4 (4\pi)}{g g^\prime} \delta \Pi^\prime_{3B}(0)  \, , \\ 
  \Delta T =& \frac{4 \pi}{g^2 s_W^2 m_W^2} \left( \delta \Pi_{+-}(0) - \delta \Pi_{33}(0)  \right)\, ,\\
  \Delta W =& -\frac12 m_W^2 \delta\Pi^{\prime\prime}_{33}(0) \, ,\\
  \Delta Y =& -\frac12 m_W^2 \delta\Pi^{\prime\prime}_{BB}(0) \, .
\end{align}
Up to second order in the mass splitting, the traces in \cref{eq:genTintSimplified,eq:genSintSimplified} reduce through the relations in \cref{app:su2conventions} to
\begin{equation}
  \Tr \left[ \left(T^3\right)^2\right] = \frac23 j (j+\frac12) (j+1) \, ,
\end{equation}
giving $\Delta S$ and $\Delta T$ to leading order in $r_1$ and $r_2$ as
\begin{align}
  \Delta S =& \frac{1}{2^\rho} \underbrace{\frac{8 }{9(4\pi)}}_{\simeq \, 0.071} \left( r_1 - \frac{2}{5} \abs{r_2}^2 \right) Y j (j+\frac12) (j+1) \, , \label{eq:generalS}\\[5pt]
  \Delta T =& \frac{1}{2^\rho} \underbrace{\frac{M^2}{9 (4\pi) s_W^2 m_W^2}}_{\simeq  \, 2.1 \left(\frac{M}{600\text{ GeV}} \right)^2} \left( r_1^2 - 4 \abs{r_2}^2\right) j (j+\frac12) (j+1) \, , \label{eq:generalT}
\end{align}
where $\rho=0 \, (1)$ for complex (real) irreps, such that the effect of a real irrep is reduced by a factor of $2$ compared to the complex equivalent. We stress that, in all cases, $\Delta S$ and $\Delta T$ effectively measure the mass splitting between Loryon components. The contributions to $\Delta W$ and $\Delta Y$ at leading order are instead independent of mass splitting,
\begin{align}
\Delta W =& \frac{1}{2^\rho} \underbrace{\frac{43}{180} \, \frac{g^2}{(4\pi)^2} \, \frac{m_W^2}{M^2}}_{\mathclap{\simeq  \, 1.2 \times 10^{-5} \left(\frac{600\text{ GeV}}{M} \right)^2}}\, \times \, j (j+\frac12) (j+1) \, , \label{eq:generalW}\\[5pt]
\Delta Y =& \frac{1}{2^\rho} \underbrace{\frac{43}{60} \, \frac{g^{\prime \, 2}}{(4\pi)^2} \, \frac{m_W^2}{M^2}}_{\mathclap{\simeq  \, 1.0 \times 10^{-5} \left(\frac{600\text{ GeV}}{M} \right)^2}}\, \times \, Y^2 (j+\frac12)\, . \label{eq:generalY}
\end{align}
For a real irrep, there is never an $r_2$ term. We have checked the above expressions with those in \cite{Zhang:2006vt} in the $r_2=0$ case, as well as with specific calculations for the 2HDM \cite{he_extra_2001, Egana-Ugrinovic:2015vgy}, a real triplet \cite{Forshaw:2001xq} and a complex triplet \cite{Cheng:2022hbo}.  

At FCC-ee, $\Delta S$ and $\Delta T$ will be measured to $\mathrm{O}(10^{-2})$ precision --- an order-of-magnitude improvement on LEP. Using the one-dimensional $2\sigma$ HL-LHC + FCC-ee bounds from \cite[Tab.\ 37]{de_blas_higgs_2020},\footnote{This assumes a central value of $\Delta S=\Delta T=0$, as well as reasonable improvements in the theory uncertainty on the SM prediction of $S$ and $T$.} \cref{eq:generalS,eq:generalT} reduce to projected constraints of
\begin{align}
  \abs{\Delta S} = \frac{1}{2^\rho} \abs{ r_1 - \frac{2}{5} \abs{r_2}^2 } \abs{Y} j (j+\frac12) (j+1) \lesssim& \, 0.22\, , \label{eq:SoneDBound}\\[5pt]
  \abs{\Delta T} = \frac{1}{2^\rho} \abs{ r_1^2 - 4 \abs{r_2}^2 } j (j+\frac12) (j+1) \lesssim& \left(\frac{600\text{ GeV}}{M} \right)^2 5.4 \times 10^{-3} \, . 
\end{align}

\begin{figure}
    \centering
    \includesvg[width=\textwidth]{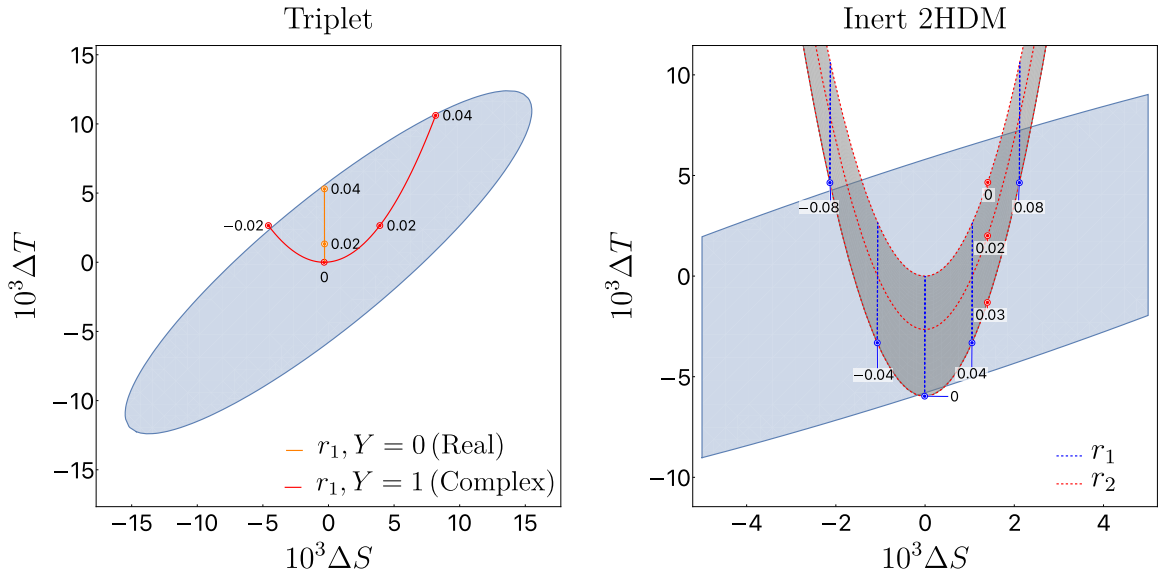}
    \caption{In blue, the projected $2 \sigma$ sensitivities in the plane of $\Delta S$ and $\Delta T$ from HL-LHC + FCC-ee \cite[Fig.\ 17]{de_blas_higgs_2020}. Left: From \cref{eq:generalS} and \cref{eq:generalT}, the $\Delta S$ and $\Delta T$ predictions as a function of the mass splitting parameter $r_1$ are shown for the real triplet (orange) and the complex triplet (red) for a common mass scale of $M$ = 600 GeV.  Right: The corresponding trajectories for the 2HDM, showing the sensitivity to both $r_1$ and $r_2$. Note that changing the common mass scale $M$ effectively rescales the $T$ axis. These plots do not include the $m_Z^2$-dependent piece in $S$ given in \cref{eq:mZDependentSpieces}.}
    \label{fig:finaloblique}
\end{figure}

In \cref{fig:finaloblique}, we show the $2 \sigma$ HL-LHC + FCC-ee projected ellipse in the two-dimensional plane of $\Delta S$ and $\Delta T$ given in \cite[Fig.\ 17]{de_blas_higgs_2020}, on top of which we show the trajectories for the 2HDM and (real and complex) triplet models obtained when varying the mass splitting parameters $r_1$ and $r_2$. For all irreps, $\Delta S$ varies linearly with $r_1$ but quadratically with $r_2$, while $\Delta T$ varies quadratically with both $r_1$ and $r_2$. Varying the common mass scale, $M$, by a few hundred GeV rescales the $\Delta T$ contribution, but does not change the order of these bounds.

Irreps with $\abs{Y}=\frac12$ explore more of the ellipse by virtue of simultaneously having $r_1$ and $r_2$ mass splittings. This is reflected in the allowed 2HDM region (shaded). For a common mass $M = 600$ GeV, and \emph{absent cancellations}, we find sensitivities for the 2HDM of the order of $\abs{r_1},\abs{r_2} \lesssim 0.1$, corresponding to mass splittings of order $10\%$. However, all $\abs{Y}=\frac12$ irreps admit a custodial limit $r_1=2\abs{r_2}$,\footnote{This can be enforced by imposing custodial symmetry. In the case of $\abs{Y}=\frac12$ irreps larger than the doublet, imposing custodial symmetry requires adding more electroweak multiplets to build full representations of the custodial symmetry group.} in which the contribution to $T$ vanishes, and $S$ alone sets a weaker bound of order $\abs{r_1} \lesssim 0.6$ for the 2HDM.

In contrast, for the real and complex triplets, only a slice of the parameter space is covered by varying $r_1$. Consequently, we find the detectable splitting between components in both the real and complex triplet is on the order of $10$ GeV.

The higher-order-in-mass-splitting terms neglected in \cref{eq:generalS,eq:generalT} are suppressed by extra powers of $r^2 j (j+1)$, where $r$ stands for $r_1$ or $r_2$. Given the loosest bound of $r \lesssim 0.6$ quoted above, $r^2 j(j+1) \lesssim 0.25$ for any irrep, and the higher-order terms are therefore subdominant.

\begin{figure}
    \centering
    \includesvg[width=0.6\textwidth]{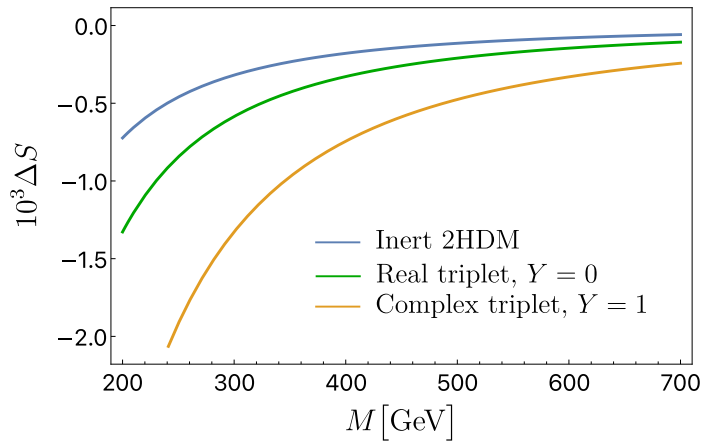}
    \caption{The $m^2_Z$-dependent shift in the $S$ parameter up to $\mathrm{O}\left(m_Z^2\right)$ for each multiplet of interest. Including this shift does not significantly change the bounds on the mass splitting parameters shown in \cref{fig:finaloblique}.}
    \label{fig:mZdependent}
\end{figure}

We note that the expressions \cref{eq:generalS,eq:generalT} are given in the $m_Z^2 \rightarrow 0$ limit and that this neglects a small $m_Z^2$-dependent shift in the $S$ parameter. Taking the zero mass-splitting limit of $S$ given in \cite{Lavoura:1993nq}, we find the shift to leading order in $m_Z^2$ for a given irrep is
\begin{equation}
  \Delta S = - \frac{1}{2^\rho} \underbrace{\frac{m_Z^2}{M^2} \frac{1}{90 \pi}}_{\mathclap{\simeq \, 8.2 \times 10^{-5} \left(\frac{600\text{ GeV}}{M} \right)^2}} \times \left( \frac12 c_W^2 j (j+\frac12) (j+1) + 6 s_W^2 Y (j+\frac12) \right) + \mathrm{O}\left(\frac{m_Z^4}{M^4}\right) \, . \label{eq:mZDependentSpieces}
\end{equation}
There is negligible difference between the all-order in $m_Z^2$ expression and the leading order piece given above. This shift is plotted in \cref{fig:mZdependent} for the smallest multiplets considered. For $M = 600 \text{ GeV}$, we find the shift in $\Delta S$ is small and makes a negligible difference to the bounds for $r_1$ and $r_2$ quoted above.

In sum, FCC-ee measurements of $\Delta S$ and $\Delta T$ are sufficient to detect any mass splitting among the components of new electroweak multiplets greater than about $10 \%$ ($60 \%$) for non-custodial (custodial) irreps, with this percentage rapidly decreasing with the dimension of the irrep.

The $W$ and $Y$ parameters probe $\mathrm{O}(p^4)$ effects in the vacuum polarisation, which are best measured in off-shell bosons at the higher centre-of-mass energy runs of FCC-ee. Using the projected 2d sensitivities for $\Delta W$ and $\Delta Y$ from \cite[footnote 16]{de_blas_higgs_2020}, we obtain from \cref{eq:generalW,eq:generalY} an approximate $2\sigma$ lower mass bound of a few hundred GeV for the smallest non-trivial electroweak multiplets: $260 \, \mathrm{GeV}$ for the 2HDM and $340 \, \mathrm{GeV}$ for the real triplet.

\subsection{Higgs Couplings: $\kappa_\gamma$, $\kappa_{h}$, $\kappa_\lambda$\label{subsec:hrun}}
\label{subsec:couplings}

In this section, we consider the effect of Loryons on the Higgs-photon coupling, the Higgs wavefunction renormalisation, and the Higgs self-coupling. These latter two are sensitive to all Loryons (charged or otherwise), and are vital in constraining the case of the neutral singlet, which proves to be the one type of Loryon that cannot be conclusively probed at FCC-ee. The results are summarised in \cref{fig:finalPlot}, which shows the sensitivities of Higgs measurements for the neutral singlet, real triplet, and inert 2HDM in the $(M,f)$ plane, and the results are discussed further in \cref{sec:discussion}.

All electrically charged particles with a Higgs coupling affect the Higgs' loop-level coupling to 2 photons. This effect is encoded in the multiplicative modifier $\kappa_\gamma$ of the coupling, which is normalised such that $\kappa_\gamma = 1$ in the SM. Using the formalism outlined in \cite{carmi_higgs_2012}, we obtain the following generic expression for $\kappa_\gamma$ in the presence of BSM charged scalars
\begin{equation}
    \kappa_\gamma = \, 1 +  \sum_{i \in \text{BSM}} \frac{ f_i \, Q^2_i}{24 \, \hat{c}_{\gamma, \text{SM}}}   A_s\left(\frac{m^2_h}{4 m^2_i}\right) \, ,
    \label{eq:kappaGamma}
\end{equation}
where the sum is over all complex BSM degrees of freedom, $\hat{c}_{\gamma,SM} = -0.81$, and
\begin{equation}
  A_s(\tau) = \frac{3}{\tau^2} \left(\arcsin^2 \sqrt{\tau} - \tau \right) = 1 + \mathrm{O}\left(\tau^2\right) \, ,
\end{equation}
assuming $\tau \equiv \frac{m^2_h}{4 m^2_i} \leq 1$.

Consider the simplest case of a singly charged scalar. Here we set $f_i \rightarrow f$ and $m_i \rightarrow M$ to express \cref{eq:kappaGamma} in terms of the notation given in \cref{eq:notation},
  \begin{align}
    \kappa_\gamma =& \, 1 +  \frac{ f }{24 \, \hat{c}_{\gamma, \text{SM}}}  A_s\left(\frac{m^2_h}{4 M^2}\right) =  1 +  \frac{f}{24 \, \hat{c}_{\gamma, \text{SM}}} + \mathrm{O}\left(\frac{m_h^2}{4 M^2}\right)^2 \, .
  \end{align}
This shows that, keeping $f$ fixed, $\kappa_\gamma-1$ is a non-zero constant in the limit of large $M$, as would be expected for a non-decoupling particle.

\begin{figure}
  \centering
  \includesvg[width=0.6\textwidth]{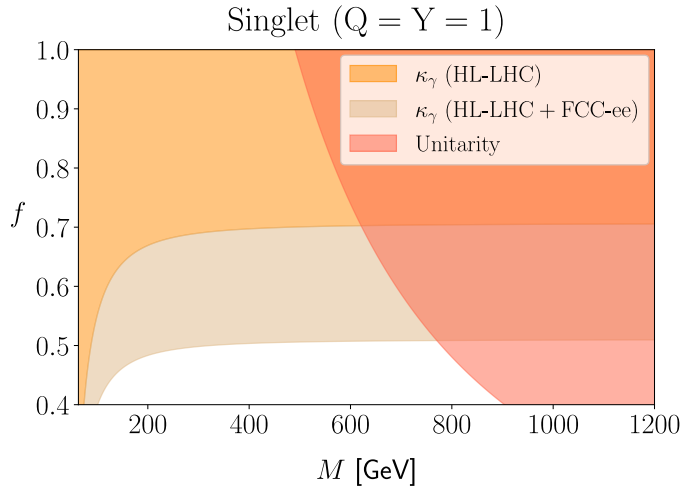}
  \caption{The parameter space for a \textit{singly charged} scalar in terms of fraction of mass-squared obtained from EWSB, $f$, \cref{eq:notation}, and total mass $M$. In red, the unitarity constraints resulting from Higgs-exchange scattering in \cref{sec:unitarity}. In orange, the projected $2 \sigma$ $\kappa_{\gamma}$ sensitivity at HL-LHC and FCC-ee. The upshot is that any charged singlets gaining the majority of their mass from EWSB (i.e., $f \geq 0.5$, and the particle exhibits non-decoupling properties, such as its SMEFT expansion not converging \cite{Cohen:2020xca}) are detectable by HL-LHC $+$ FCC-ee.}
  \label{fig:kappa_gamma}
\end{figure}

Ref.\ \cite[Tab.\ 4.2]{abada_fcc_2019} gives projected $2\sigma$ sensitivities to $\kappa_\gamma$ of $3.6\%$ at HL-LHC and $2.6\%$ using HL-LHC+FCC-ee, including runs at $\sqrt{s}$ of both 240 GeV and 365 GeV. In \cref{fig:kappa_gamma}, we show that any charged scalar that fits the $\textit{Loryon}$ description, for which $f \geq 0.5$, could therefore be ruled out. Note that $\kappa_{Z\gamma}$ gives a linearly-independent, but significantly less sensitive bound \cite{banta_non-decoupling_2022}.

While $\kappa_{\gamma}$ plays a significant role in constraining any electrically charged states, at the FCC-ee a complementary constraint comes from Higgs wavefunction renormalisation $\kappa_h$ \cite{Englert:2013tya, Katz:2014bha, Curtin:2014jma, Huang:2016cjm}. This effect exists regardless of the electroweak charges of the multiplet, and so is vital in constraining the case of the scalar singlet.

$\kappa_h$ describes a finite correction to the kinetic term of the Higgs from a loop of heavy particles. This translates into a uniform shift of all single Higgs couplings once the kinetic term is canonically normalised
\begin{equation}
    \lag = \frac12 \kappa_h \left( \partial h\right)^2 - h \, J \longrightarrow \frac12 \left( \partial h\right)^2 - \kappa_h^{-\frac12} h \, J \, ,
\end{equation}
where $J$ denotes all scalar currents coupling to the Higgs. If the heavy particles do not have SM charges, the products of Higgs cross-sections times branching ratios all scale uniformly as $\kappa_h^{-1}$ at one-loop. For a fractional uncertainty $\sigma_i$ on each cross-section times branching ratio, this leads to an effective uncertainty $\sigma_\text{eff}$ on $\kappa_h^{-1}$ of approximately $0.23\%$
\begin{equation}
  \sigma_\text{eff} = \left( \sum_{i \in \text{modes}} \frac{1}{\sigma_i^2} \right)^{-\frac12} \approx 0.0023 \, , \label{eq:sigmaEff}
\end{equation}
where we have taken the $\sigma_i$ for the $\sqrt{s} = 240$ GeV, 5 $\text{ab}^{-1}$ run from \cite[Tab.\ 4.1]{abada_fcc_2019}.

The contribution of the general multiplet to $\kappa_h$ follows from its contribution to the Higgs vacuum polarisation, which we define by the interaction $\lag = \frac12 \Pi_{hh}(p^2) h(p) h(-p)$.

In the notation of \cref{sec:genericLag}, the multiplet contributes
  \begin{align}
    2 \left(4 \pi\right)^\frac{d}{2} \delta \Pi_{hh}^\prime(0) =& \frac{1}{\Gamma\left({\textstyle \frac{d}{2}}+1\right)} \int \dd t \, t^{\frac{d}{2}} \Tr \left[ \Delta^{-2} \left(\mK^{-1} \mM^\prime\right) \Delta^{-2} \left(\mK^{-1} \mM^\prime\right) \right] \, ,
  \end{align}
in terms of the partial derivative of $\mM$ with respect to $v$, $\mM^\prime$. Expanding to second order in the mass splitting, this reduces to
\begin{align}
  \delta \Pi_{hh}^\prime(0) =& \, \kappa_h-1 \, , \notag\\
  =& \frac{1}{2^\rho} \underbrace{\frac{2M^2}{3\left(4 \pi\right)^2 v^2}}_{\mathclap{\approx 0.025\left(\frac{M}{600\text{ GeV}} \right)^2}} (2j+1) \left( f^2  + \frac13 j(j+1)\, (1-f)^2 (r_1^2 - 8 \abs{r_2}^2) \right) + \mathrm{O}\left(r^3\right) \, . \label{eq:kappahCorrection}
\end{align}
Coincidentally, the expected deviation due to one real Loryon degree of freedom is comparable to the expected experimental sensitivity.

Analogous to the $m_Z^2$ dependence of $S$, by evaluating the vacuum polarisation at zero momentum, \cref{eq:kappahCorrection} neglects an $m_h^2$ dependence which at leading order is
\begin{equation}
  \kappa_h-1 = \frac{1}{2^\rho} \underbrace{2 \frac{m_h^2}{15\left(4 \pi\right)^2 v^2}}_{\mathclap{\approx 2.2 \times 10^{-4}}} \times (2j+1) f^2  \, ,
\end{equation}
in the absence of mass splitting. This correction is too small to be relevant for our analysis.

The size of $\sigma_\text{eff}$, \cref{eq:sigmaEff}, is controlled by the most precisely constrained cross-section times branching ratio: that of $h \to ZZ$. If the multiplet has electroweak charges, the $h \to ZZ$ rate also receives a direct contribution from a loop of heavy particles connected to the $h$ and $Z$ bosons, in addition to the contribution from wavefunction renormalisation of the $h$. However, for a particle whose Higgs coupling is larger than its gauge coupling squared, the direct contribution generically results in a smaller fractional shift in the measured rate. The size of this direct contribution is given at leading order by the insertion of an effective operator
\begin{equation}
  \lag_\text{eff} = c_{ZZ} \frac{h}{v} Z_{\mu\nu} Z^{\mu\nu} \, ,
\end{equation}
where $c_{ZZ}$ is calculated in the high mass limit analogously to the $\kappa_\gamma$ calculation above, giving
\begin{equation}
  c_{ZZ} = \frac{2}{3 (4 \pi)^2} f \left( \frac23 g^2 c_W^2 j (j+\frac12) (j+1)  + g_1^2 s_W^2 (2j+1) Y^2\right) \, .
\end{equation}
Using the phase space factor reported in \cite{Boselli:2017pef}, the effective operator results in a fractional shift in the $h \to ZZ$ partial width of
\begin{equation}
  \frac{\delta \Gamma}{\Gamma} = - 2.1 \times 10^{-4} \times f j (j+\frac12) (j+1) - 2.9 \times 10^{-5} \times f (2j+1) Y^2 \, .
\end{equation}
This effect is an order of magnitude smaller than the per mille level precision achieved at FCC-ee. Thus, we expect that it is still a reasonable approximation to bound \cref{eq:kappahCorrection} using the experimental sensitivity \cref{eq:sigmaEff}, even when the multiplet has electroweak charges.

For comparison, we also quote from \cite{banta_non-decoupling_2022} the effect on the Higgs trilinear
\begin{equation}
  \label{eq:kappalambda}
  \kappa_{\lambda} -1 = \frac{1}{2^\rho} \underbrace{\frac{16}{3\,(4\pi)^2} \frac{M^4}{m^2_h v^2}}_{\approx 2.3 \left(\frac{M}{600\text{ GeV}} \right)^4} f^3 \left (2\,j+1 \right ) \, ,
\end{equation}
on which we conservatively expect $2\sigma$ sensitivities of order $100\%$ at HL-LHC, and $10\%$ at FCC-hh \cite{abada_fcc_2019}.

The projected sensitivities of $\kappa_\gamma$, $\kappa_h$, and $\kappa_\lambda$ are shown for the parameter space of the smallest electroweak irreps in \cref{fig:finalPlot}. Important for FCC-ee is that the sensitivity on $\kappa_h$ surpasses the unitarity bound on the Higgs coupling, meaning that the measurement of $\kappa_h$ will be a useful indirect measurement of any new particle's Higgs coupling, independently of said new particle's gauge charges. For a new scalar getting all of its mass from the Higgs, this constitutes a $2\sigma$ sensitivity to masses above $400$, $220$ and $190$ GeV for the neutral singlet, real triplet, and 2HDM respectively. In the case of all models but the neutral singlet, this is as good as the sensitivity arising from the loop-level contribution to $\kappa_\lambda$ at FCC-hh.

\section{Electroweak Phase Transition} 
\label{sec:baryogenesis}

\newcommand{\hsym}{\mathfrak{h}}

TeV scale particles with a strong coupling to the Higgs can modify the electroweak phase transition that likely occurred in the early universe. This is a vital part of a theory of electroweak baryogenesis which explains the outstanding problem of baryon asymmetry \cite{Kuzmin:1985mm, Shaposhnikov:1987tw} (see \cite{Morrissey:2012db} for a review), as the phase transition can satisfy the necessary Sakharov condition of departure from thermal equilibrium \cite{Sakharov:1967dj}. 

A theory supporting electroweak baryogenesis requires the potential to allow a strongly first order phase transition (SFOPT). In this scenario, a barrier forms between the symmetry breaking and symmetry preserving minima in the potential at temperatures of $\mathrm{O}(100 \text{ GeV})$. Tunnelling between the minima allows bubbles of the new symmetry breaking phase to nucleate throughout the Universe.

Studying the SM potential, a SFOPT is not observed. In terms of the field $\hsym=v+h$, there is instead a smooth crossover transition between the symmetry preserving phase centred around $\langle \hsym \rangle = 0$ and the symmetry breaking phase $\langle \hsym \rangle = v$. However, a potential barrier is induced via additional contributions to the potential from BSM scalars, whose effect is to enhance the $\hsym^3$ Higgs cubic term in the finite temperature potential. If a SFOPT is initiated, bubble nucleation of the new phase marks the departure from thermal equilibrium and can initiate CP violating processes at the bubble wall, which convert into a baryon asymmetry through sphaleron transitions in the symmetry preserving phase. As baryons enter the bubble, the sphaleron transition rate is suppressed by EWSB, ensuring the baryon asymmetry stabilises.

Nucleation of bubbles of the new phase can generate gravitational waves (GWs) via collisions between bubble walls, sound waves as well as plasma turbulence (see \cite{Croon:2023zay} for review). The resulting stochastic background of gravitational waves is a target of the next generation of detectors such as the LISA telescope. 

Here we follow the work of \cite{banta_strongly_2022} in analysing the scalar Loryon parameter space that admits a SFOPT and generates a stochastic gravitational-wave background, to which we refer the reader for more details.

\subsection{Strong First Order Phase Transitions}

The masses of the ensemble of particles change during the phase transition due to their dependence on the value of the Higgs field. Denoting the field dependent masses as $m^2_i(\hsym)$, we use the field dependent masses for SM particles as defined in \cite{Noble:2007kk,Croon:2023zay} and convert the scalar masses defined in \cref{tab:modelMassSpectra} to be field dependent through $v \rightarrow \hsym \equiv v + h$, noting that the parameter $M^2$ depends implicitly on the Higgs vev, and therefore also acquires $\hsym$ dependence. By construction, the Loryons do not get a vev at any point during the phase transition, so the effective potential, denoted $V_{\mathrm{eff}} (\hsym)$, is only dependent on $\hsym$.

The effective potential includes a zero temperature piece consisting of the tree-level potential $V_\mathrm{0}$ and one-loop Coleman-Weinberg potential $V_\mathrm{CW}$ \cite{Coleman:1973jx}. Additionally, there are finite temperature corrections $V_\mathrm{T}$. For our ensemble of bosons (bos) and fermions (fer), the full effective potential reads,
  \begin{multline}
  \label{effective_potential}
     V_{\mathrm{eff}} (\hsym) = V_{0}(\hsym) + \underbrace{\sum_{i \in \text{SM}} n_i V_{\mathrm{CW,bos}}(m_{i}^{2}(\hsym)) + n_t V_{\mathrm{CW,fer}}(m_{t}^{2}(\hsym)) + n_{\Phi} V_{\mathrm{CW,bos}}(m_{\Phi}^{2}(\hsym))}_{\text{zero temperature corrections}} \ \\
     +  \underbrace{\sum_{i \in \text{SM}} n_i V_{\mathrm{T,bos}}(m_{i}^{2}(\hsym),T) +  n_t V_{\mathrm{T,fer}}(m_{t}^{2}(\hsym),T) + n_{\Phi} V_{\mathrm{T,bos}}(m_{\Phi}^{2}(\hsym),T) \, ,}_{\text{finite temperature corrections}} \
    \end{multline}
where the bosonic SM contribution comes from $i$ = \{$W_T, W_L, Z_T, Z_L, h, \chi, \gamma_L$\} and the corresponding real degrees of freedom for the SM boson ensemble are $n_i$ = \{4, 2, 2, 1, 1, 3, 1\}. The top quark $t$ is the only fermion considered, with $n_t=12$, as the others are suppressed by small Higgs couplings. We include the contributions from Loryons $\Phi$, for which the degrees of freedom for each multiplet can be found in \cref{tab:modelMassSpectra}. We work in the Landau gauge which requires the inclusion of the Goldstone modes $\chi$. This breaks gauge invariance; however, as discussed in, e.g., \cite{Carena:2019une}, the gauge dependence is found at loop level whereas the strongest effects on the phase transition are at tree level, so this does not compromise our analysis significantly.

Expressions for $V_\mathrm{CW}$ and $V_\mathrm{T}$ are given to one loop in \cref{effective_terms}. There is an additional mass shift of the field dependent masses due to daisy resummation of thermal loops \cite{Carrington:1991hz}. We use the Parwani scheme, inserting $m_{i}^{2}(\hsym) \rightarrow m_{i}^{2}(\hsym) + \Pi_i(\hsym,T)$ into \cref{effective_potential} \cite{Parwani:1991gq}, where the finite temperature mass corrections for all particles in our ensemble are evaluated in the high temperature limit, and given in \cref{daisy_terms}. Note that the SM Higgs and the Goldstones mass-squareds are negative for small values of $\hsym$, making the one-loop contributions to the potential complex. For phase transitions only the real part of the potential need be considered, as this is the dominant contribution \cite{Delaunay:2007wb,Alonso:2023jsi}.

For a first order phase transition to be strong enough for baryogenesis, a constraint of $100 < S_3/T_{\mathrm{nuc}} < 200$ is imposed \cite{McLerran:1990zh, Anderson:1991zb}. As in \cite{banta_strongly_2022}, this range intended to bracket the desired value of $S_3/T_{\mathrm{nuc}} \approx 140$ and account for potential uncertainties in our calculation of $S_3/T_{\mathrm{nuc}}$. Here $T_{\mathrm{nuc}}$ is the nucleation temperature, marking the onset of the phase transition, defined as the temperature at which a bubble of the new phase nucleates in one Hubble time. The bounce action $S_3$ of the instanton controls the rate of the tunnelling from one vacuum to another \cite{Quiros:1999jp}. Both of the above parameters are found through the \texttt{CosmoTransitions} package \cite{Wainwright:2011kj} using our construction of the effective potential as detailed above. Moreover, we require $v_{\text{nuc}}/T_{\text{nuc}} > 1$, where $v_{\text{nuc}}$ is the symmetry breaking vev at the nucleation temperature, ensuring that sphaleron transitions are sufficiently suppressed inside the bubble \cite{Quiros:1999jp}. We note that including higher-loop order effects in the thermal potential can enhance $v_{\text{nuc}}/T_{\text{nuc}}$, and thus loosen this constraint \cite{Cohen:2011ap}. The viable regions for a SFOPT are shown in the limit of no mass splitting for the smallest irreps in \cref{fig:finalPlot}.

In \cref{fig:finalPlot}, the lower left bound is given by the largest possible $T_{\text{nuc}}$ that satisfies both 100 $ \lesssim S_3/T_{\mathrm{nuc}} \lesssim$ 200 and $v_{\text{nuc}}/T_{\text{nuc}} > 1$. The upper right bound is given by the lowest possible $T_{\text{nuc}}$. In practice, we take this lowest value to be 50 GeV, as we find that scanning any lower does not visibly shift the bound, in agreement with \cite{banta_strongly_2022}.

We also explored the effect of the mass splitting parameters $r_1$ and $r_2$ on the phase transition; however, this is minimal when enforcing the bounds on $r_1$ and $r_2$ that would result from the $Z/WW$ runs of FCC-ee (see \cref{subsec:zwwrun}). Their lack of noticeable effect is partly due to the Coleman-Weinberg and thermal potentials being quadratic at leading order in $r_1$ and $r_2$. This is because $r_1$ and $r_2$ give equal and opposite shifts in the masses of different components of multiplet (see examples in \cref{tab:modelMassSpectra}); the potentials sum over the same function of the mass for all components, meaning that the linear pieces in $r_1$ and $r_2$ vanish. This also means that neither parameter has an influence on the leading-order thermal mass corrections, as shown by the lack of $\lambda^{\prime}$ or $\lambda^{\prime\prime}$ terms in \cref{tab:daisy}.

\subsection{Gravitational Wave Background}

The strength of the gravitational wave background can be characterised by the thermal energy released during the transition, as well as its duration. The parameter $\alpha$ is defined as the vacuum energy density expelled during the phase transition compared to the total radiation energy density,
\begin{equation}
\label{eq:alpha}
    \alpha = \left (\Delta V_{\mathrm{eff}} - \frac{T_{\mathrm{nuc}}}{4} \Delta \frac{\dd V_{\mathrm{eff}} }{\dd T} \right ) \Big/ \frac{g_{\mathrm{eff}}(T)\, \pi^2\, T^4_{\mathrm{nuc}}}{30},
\end{equation}
where $\Delta$ indicates the difference between the symmetric and broken phases, and $g_\mathrm{eff} (T)$ = $g_\mathrm{SM} (T)$ +  $g_\mathrm{BSM} (T)$ are the effective relativistic degrees of freedom of the particle plasma in the symmetric phase. For all electroweak multiplets considered, we fix $g_\mathrm{eff} (T)$ = 100, introducing a sub-dominant uncertainty $\sim \,10 \%$ (a more precise calculation should utilise the functional form of $g_\mathrm{eff} (T)$ found in \cite{Laine:2015kra}).

The second parameter of interest is the inverse duration of the phase transition, $\beta$, which takes the form,
\begin{equation}
\label{eq:beta}
    \beta / H_{*} = \frac{\dd S_\mathrm{3}}{\dd T} \bigg|_{T_{\mathrm{nuc}}} - \frac{S_\mathrm{3}}{T_{\mathrm{nuc}}} \, ,
\end{equation}
where $H_{*}$ denotes the Hubble time. Combined, the parameters $\alpha$ and $\beta$ provide a constraint on the phase transitions that could produce a residual GW background potentially detectable by the next-generation GW detectors. For the LISA telescope, these bounds are given by \cite{banta_strongly_2022}
\begin{equation}
\label{eq:GW_constraints}
    \mathrm{log}(\beta / H_{*}) - 1.2 \ \mathrm{log}(\alpha) < 8.8 \, .
\end{equation}
The points satisfying this bound are shown in \cref{fig:finalPlot}. The allowed region for a detectable GW signal closely follows the upper right boundary of the SFOPT region, which indicates that SFOPTs occurring at lower nucleation temperatures produce stronger GW signals. As shown in \cref{eq:alpha,eq:beta}, this corresponds to a shorter phase transition with more energy released.

We stress again that the above analysis contains many approximations, and the boundaries of the SFOPT and GW regions drawn in this section may move in a fuller analysis. However, this calculation should be sufficient to identify the approximate region of interest in the $(M,f)$ plane.

\section{Results and discussion\label{sec:discussion}}

\begin{figure}
\centering
\includesvg[width=\textwidth]{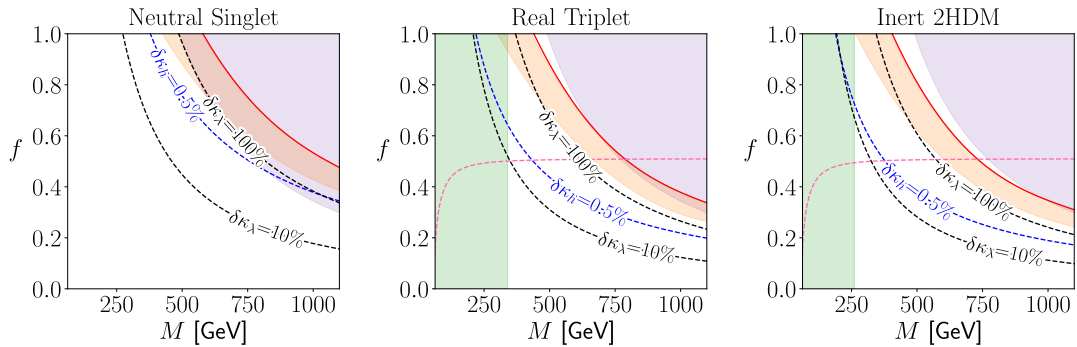}
\caption{The parameter space of the three smallest multiplets in the absence of mass splitting, shown as the fraction of mass gained from EWSB, $f$, against total mass $M$. The purple region is excluded by unitarity constraints on the multiplet's Higgs coupling (see \cref{sec:unitarity}). The green region is probed by FCC-ee $2\sigma$ constraints on $\Delta W$ and $\Delta Y$ (see \cref{subsec:zwwrun}). Regions in which a strongly first order phase transition can occur are denoted by the orange band. The red line denotes the lower bound for a detectable GW signal in LISA, given by \cref{eq:GW_constraints}. The regions above dashed lines are detectable by future Higgs measurements. An approximate $0.5\%$ $2\sigma$ FCC-ee sensitivity to $\kappa_h$ is shown by the dashed blue line (see \cref{eq:kappahCorrection}). Approximate future sensitivities for $\kappa_{\lambda}$ are given in the black dashed lines for the HL-LHC ($100\%$) and FCC-hh ($10\%$) (see \cref{eq:kappalambda}). Projected sensitivity to $\kappa_{\gamma}$ at FCC-ee is shown as the pink line for charged multiplets (see \cref{eq:kappaGamma}).}
\label{fig:finalPlot}
\end{figure}

In this Section, we consider how the combination of unitarity constraints (\cref{sec:unitarity}), future electroweak (\cref{subsec:zwwrun}) and Higgs (\cref{subsec:hrun}) measurements, and SPOFT and GW regions (\cref{sec:baryogenesis}) behave in the non-decoupling parameter space of electroweak scalar multiplets.

Assume there exists a BSM electroweakly-charged multiplet. Once the possibility of tree-level mixing with SM particles is discounted, the effect of the $Z/WW$ runs at a future $e^+e^-$-collider is to detect any mass splitting between the components of the multiplet via the components' loop-level oblique effects in $S$ and $T$. We project a sensitivity of a few tens of GeV of mass splitting for a TeV scale multiplet, loosening to a couple of hundred GeV for a custodial multiplet. Note that there is less than an order of magnitude difference between the custodially symmetric and violating cases. In the absence of tree-level mixing with the gauge bosons, future electroweak precision tests will not vastly prefer custodially symmetric over custodially violating physics: they will prefer small mass splittings among components of electroweak multiplets. As shown in \cref{subsec:zwwrun}, after the $Z/WW$ runs of FCC-ee, the assumption of no mass splitting for undiscovered electroweak multiplets will be well justified. We make this assumption henceforth.

We consider this no-mass-splitting scenario in \cref{fig:finalPlot} for the neutral singlet, real triplet, and inert 2HDM models. Theoretical bounds come from unitarity constraints outlined in \cref{sec:unitarity}, where in the absence of mass splitting the constraints are the same for each model.

$\Delta W$ and $\Delta Y$ give an $f$-independent probe of the mass of electroweakly charged multiplets, shown by the green band in \cref{fig:finalPlot}. However, for each multiplet, there is a region in the parameter space where a SFOPT is possible, and which is not excluded by these electroweak or unitarity constraints. In the case of the singlet, a possible SFOPT can occur around $f \gtrsim 0.65$; however, the region for a viable resulting gravitational wave background is ruled out by unitarity constraints. The SFOPT and GW regions for the 2HDM and triplet are shifted downwards due to a larger number of degrees of freedom compared to the singlet. The effect of this is to broaden the viable region for a SFOPT and introduce a slice of the parameter space where a stochastic GW background could be detected.

As shown by the blue dashed line in \cref{fig:finalPlot}, on-shell Higgs measurements at FCC-ee are sensitive to the entire SFOPT region. Any scalar with large Higgs couplings (that are sufficient to modify the order of the phase transition) is visible via the electroweak-charge-independent measurement of the Higgs wavefunction normalisation, $\kappa_h$, using the $2 \sigma$ projected sensitivity to $\kappa_h$ of $0.5\%$. Note that a more conservative precision of $1\%$ or $2\%$ would be inadequate to probe the desired parameter space. For each multiplet in \cref{fig:finalPlot}, the projected $\kappa_h$ sensitivity lies between the $\kappa_\lambda$ sensitivity of $100\%$ at HL-LHC and $10\%$ at FCC-hh.

Furthermore, now including $\kappa_\gamma$ (the pink line in \cref{fig:finalPlot}), Higgs measurements at FCC-ee are sensitive to any non-decoupling electroweakly-charged multiplets that get the majority of their mass from EWSB (i.e, with $f > \frac12$). The only invisible non-decoupling scenario would be a $\mathbb{Z}_2$-symmetric real scalar singlet, which on its own is insufficient to cause a SFOPT. For the singlet, $\kappa_h$ at FCC-ee makes inroads, as well as $\kappa_\lambda$ at FCC-hh which does even better, but neither can see, e.g., a $200$ GeV mass singlet, even if all of its mass comes from a coupling to the Higgs. We note here that contributions to $\kappa_h$ and $\kappa_\lambda$ scale as the number of degrees of freedom,\footnote{The finite temperature effective potential, and therefore also the SFOPT and GW regions in \cref{fig:finalPlot}, also primarily depend on the number of degrees of freedom for fixed $M$ and $f$.} so the existence of several singlet scalars could be detected (cf.\ the bounds for $3$ and $4$ real degrees of freedom in the real triplet and 2HDM case). Contributions to $\kappa_h$ from scalars are all the same sign, and cannot cancel.\footnote{Fermions give opposite sign contributions to $\kappa_h$; however, a sizeable contribution would imply a sizeable renormalisable Yukawa coupling, and in turn electroweak charges for the corresponding fermion multiplet, which should be detectable via $\kappa_\gamma$, $\kappa_{Z\gamma}$ at HL-LHC \cite{banta_non-decoupling_2022,Barducci:2023zml}.}

\section{Conclusions\label{sec:conclusions}}

A scalar multiplet with significant Higgs couplings is a potentially viable model of new physics that significantly alters EWSB. In this paper we considered its minimal loop-level effects in future precision observables.

We have shown that the $Z$, $WW$, and $h$ runs of FCC-ee are sensitive to any scalar multiplet with electroweak charges whose components obtain the majority of their mass from interactions with the Higgs. Thus, FCC-ee is generically sensitive to an interesting class of non-decoupling physics, which significantly changes the nature of the electroweak phase transition, and which also predicts a pattern of effects in low energy observables that is not well characterised by the power counting in SMEFT. This generic exercise also serves to reemphasise the complementarity between observations of the stochastic gravitational wave background generated by a putative first order electroweak phase transition in the early universe, and future precision Higgs measurements.

As similar non-decoupling coloured particles and fermions contain more degrees of freedom and could likely be seen at HL-LHC already \cite{banta_non-decoupling_2022,Barducci:2023zml}, this is tantamount to saying that, absent a conspiracy of cancellations, \emph{all particles} that obtain the majority of their mass from the Higgs could be seen by the end of the FCC-ee run, except for one or two $\mathbb{Z}_2$-symmetric real scalar singlets with mass of a few hundred GeV. This general claim is achieved primarily by looking for deviations in universal two-point functions, not just for the electroweak gauge bosons, but also now for the Higgs boson itself. Although we have used FCC-ee as a benchmark here, similar conclusions can be drawn for any lepton collider capable of producing $10^6$ Higgses in a clean environment.

As the nature of electroweak symmetry breaking is of fundamental importance, and as the non-decoupling physics that affects it has a finite and probeable target space, we believe it forms an important part of the physics case of such a future collider.

\acknowledgments

DS wishes to thank I.\ Banta, T.\ Cohen, N.\ Craig, and X.\ Lu for collaboration on the earlier related work \cite{banta_non-decoupling_2022}. We are especially indebted to I.\ Banta for help in reproducing the constraints of \cite{banta_strongly_2022}. We thank T.\ Cohen and S.\ Renner for detailed comments on the manuscript. GC is supported by STFC grant ST/Y509188/1. DS is supported by STFC grant ST/X000605/1.

\appendix
\begin{appendices}

\section{$SU(2)$ conventions\label{app:su2conventions}}

  For generators in a complex rep
  \begin{equation}
    \left(T^i\right)^\dagger = T^i \, ; \qquad \left[T^i,T^j\right] = i \epsilon^{ijk} T^k \, .
  \end{equation}
  We choose to write them in a basis where $T^3$ is diagonal, $T^1$ is real, and $T^2$ is imaginary. In the dimension-$2j+1$ irrep, letting $\alpha,\beta=j,j-1,\ldots,-j$, the generators are
  \begin{align}
    (T^1)_{\alpha\beta} \pm i (T^2)_{\alpha\beta} = (T^\pm)_{\alpha\beta} =& \sqrt{j(j+1)-\beta(\beta\pm 1)} \, \delta_{\alpha \mp 1, \beta} \, ,\\
    (T^3)_{\alpha\beta} =& \beta \, \delta_{\alpha \beta} \, .
  \end{align}

  For generators in a real irrep (only possible when $j$ integer)
  \begin{equation}
    \left(T^i_R\right)^T = -T^i_R \, ; \qquad \left[T_R^i,T_R^j\right] = \epsilon^{ijk} T_R^k \, .
  \end{equation}
  In terms of the transformation
  \begin{equation}
    M_{\alpha\beta} = \begin{cases}
      1 & \alpha=\beta=0 \\
      \frac{(-1)^\beta}{\sqrt{2}} & \alpha=\beta>0 \\
      \frac{1}{\sqrt{2}} & \alpha=-\beta>0 \\
      \frac{i}{\sqrt{2}} & \alpha=\beta<0 \\
      \frac{-i(-1)^\beta}{\sqrt{2}} & \alpha=-\beta<0 \\
      0 & \text{otherwise}
    \end{cases} \, ,
  \end{equation}
  and the complex irrep, we work in the basis where
  \begin{equation}
    T_R^i = -i M T^i M^{-1} \, .
  \end{equation}

  Using our conventions for the irreps, we can define
  \begin{equation}
    S_{\alpha\beta} = (-1)^{j-\beta-1} \delta_{-\alpha,\beta} \, , 
  \end{equation}
  which satisfies
  \begin{equation}
    S = S^* = S^{-1} \, ,
  \end{equation}
  making tilded versions of any irrep $\Phi$
  \begin{align}
    \tilde{\Phi}_\alpha =& S_{\alpha\beta} \Phi^*_\beta \, .
  \end{align}
  This maps irreps to their (equivalent) conjugate irrep, and satisfies
  \begin{equation}
    -S \left(T^i\right)^* S^{-1} = T^i \, .
  \end{equation}

\section{Effective Potential Terms}
\label{effective_terms}

This Appendix presents expressions for the Coleman-Weinberg and one-loop thermal contributions to the effective potential used in \cref{effective_potential}. 

Here, we write $V_{\text{CW}}$ in terms of the $\hsym$-dependent masses, $m^{2}(\hsym)$, as well as the masses evaluated at the $T=0$ vev, $m^{2}(v)$. For the Loryon mass spectrum given in \cref{tab:modelMassSpectra}, field dependence arises by converting  $v \rightarrow \hsym$, noting the overall mass scale, $M$, will have the following field dependence, $M^2 (\hsym) = m_\text{ex}^2 + \frac12 \lambda \hsym^2$. For bosons and fermions, we use the definition,
\begin{equation}
    V_{\mathrm{CW},\mathrm{bos/fer}}(m^{2}(\hsym)) = \pm \frac{1}{64 \pi^2}  m^{2}(\hsym) \left( m^{2}(\hsym) \left(  \log \frac{m^{2}(\hsym)}{m^{2}(v)} - \frac{3}{2} \right) + 2 m^{2}(v) \right).
    \label{coleman_weinberg}
\end{equation}
The potential is given in the on-shell scheme, which fixes the counterterm contributions for bosons and fermions and ensures that the following conditions are satisfied,
\begin{equation}
\label{renormalisable}
    \frac{\dd V_{\mathrm{eff}, T\mathrm{=0}}}{\dd \hsym} \bigg|_{\hsym=v} = 0, \qquad \frac{\dd^2 V_{\mathrm{eff}, T\mathrm{=0}}}{\dd \hsym^2} \bigg|_{\hsym=v} = m^2_{h} \, .
\end{equation}

For $V_{\mathrm{T}}$, we use the integral,
\begin{equation}
   J_{\text{bos/fer}}(y^2) =  \pm \int_0^{\infty} dx  \ x^2 \ \mathrm{log}\left(1 \mp \exp(-\sqrt{x^2 + y^2})\right) \, ,
\end{equation}
allowing $V_{T}$ at one-loop to be expressed as
\begin{equation}
    V_{T,\mathrm{bos/fer}}(m^{2}(\hsym)) \approx \frac{T^4}{2 \pi^2} J_{\text{bos/fer}}\left(\frac{ m^{2}(\hsym)}{T^2}\right) \, .
    \label{one_loop}
\end{equation}
In the literature, analytic approximations of the integral $J_{\text{bos/fer}}(y^2)$ are often used for the high temperature limit $y \ll 1 $ or the low temperature approximation $y \gg 1 $ \cite{Hindmarsh:2020hop, Curtin:2016urg}. Neither is appropriate for our purposes, as we consider a mass and temperature range that spans from $\mathcal{O}(1) \leq y \leq \mathcal{O}(10)$. We numerically evaluate the exact integral.

\section{Daisy Resummation Terms}
\label{daisy_terms}

The field-dependent masses of boson $i$ are shifted by an amount $\Pi_i$ due to resummation of hard thermal loops, which we report here at leading order in the high temperature expansion. 

For every real scalar field $f_i$ in the theory, we calculate the shift from the thermal potential in the high temperature limit to be
\begin{equation}
  \Pi_{i } = \frac{\partial^2 V_T}{\partial f_i^2} \, .
\end{equation}
For the SM Higgs and Goldstones 
\begin{equation}
  \Pi_h = \Pi_\chi = \frac{1}{24} T^2 \left(\frac{3}{2} {g^\prime}^2 +\frac{9}{2} g^2 + 12 \lambda_{hh} + 6 y_t^2 + n_\text{Loryons} \lambda \right) \, ,
\end{equation}
where the Higgs self-coupling is $\lambda_{hh}$ and $n_\text{Loryons}$ is the number of real Loryon degrees of freedom. The shifts in the Loryon masses are given in \cref{tab:daisy}.

For the SM gauge bosons, only the longitudinal modes acquire thermal corrections as the corresponding corrections to the transverse modes are suppressed due to gauge symmetry \cite{Espinosa:1992kf}. We rescale the contribution of the Higgs piece calculated in \cite{Carrington:1991hz,Seller:2023xkn} to arrive at the overall SM+Loryon expressions
\begin{align}
  \Pi_{W_\mathrm{L}} =& \frac{11}{6} g^2 T^2 + \frac{1}{2^\rho} \frac{2}{9} j(j+\frac12)(j+1) g^2 T^2 \, ,\\
  \Pi_{B_\mathrm{L}} =& \frac{11}{6} {g^\prime}^2 T^2 + \frac{1}{2^\rho} \frac{1}{3} Y^2 (2 j+1) {g^\prime}^2 T^2 \, ,
\end{align}
which we report for our models of interest in \cref{tab:daisy}. The thermal corrected masses $m_{Z_{\mathrm{L}}}^{2} + \Pi_{Z_{\mathrm{L}}}$ and $m_{\gamma_\mathrm{L}}^{2} + \Pi_{\gamma_\mathrm{L}}$ are therefore given by the appropriate eigenvalues of the matrix \cite{Beniwal:2017eik}
\begin{equation} 
\begin{pmatrix}
  \frac{g^{2}\hsym^2}{4} + \Pi_{W_\mathrm{L}} & -\frac{{g^\prime} g \hsym^2}{4} \\ -\frac{{g^\prime} g \hsym^2}{4} &  \frac{{g^\prime}^{2}\hsym^2}{4} + \Pi_{B_\mathrm{L}}
\end{pmatrix} \, .
\end{equation}

\begin{table}
\renewcommand{\arraystretch}{1.5}
\centering
\begin{tabular}{cccc}

 & \textbf{Singlet}&\textbf{Real triplet}&\textbf{Inert 2HDM}\\
\cmidrule{2-4}\morecmidrules\cmidrule{2-4}
$n_\text{Loryons}$&$1$& $3$ & $4$ \\ 
$\Pi_{W_{L}}$&$\frac{11}{6} g^2 T^2$& $\frac{13 g^{2}T^2}{6}$ & $\frac{12 g^{2}T^2}{6}$ \\ 
$\Pi_{B_{L}}$&$\frac{11}{6} {g^\prime}^2 T^2$& $\frac{11 {g^\prime}^{2}T^2}{6}$ & $\frac{12 {g^\prime}^{2}T^2}{6}$ \\ 
$\Pi_{S}$&$\frac{1}{24} T^2 \left(4 \lambda \right)$& &\\ 
$\Pi_{H^0} = \Pi_{H^\pm}$&&$ \frac{1}{24} T^2 \left(12 g^2 + 4 \lambda \right) $&\\ 
$\Pi_H = \Pi_A = \Pi_{H^\pm}$& & & $\frac{1}{24} T^2 \left(\frac{3}{2} {g^\prime}^2 +\frac{9}{2} g^2 + 4 \lambda \right)$\\

\end{tabular}
\caption{Daisy corrections from the resummation of hard thermal loops for the longitudinal gauge bosons and the scalar components that comprise the neutral singlet, real triplet and inert 2HDM.
\label{tab:daisy}}
\end{table}

\end{appendices}
\newpage

\bibliographystyle{JHEP}
\bibliography{Refs.bib}

\end{document}